**PAPER • OPEN ACCESS**

# The use of multi-agent systems for modeling technological processes

View the article online for updates and enhancements.

# The use of multi-agent systems for modeling technological processes

**Sergey Petrovich Bobkov and Irina Aleksandrovna Astrakhantseva**[0000-0003-2841-8639]

Ivanovo State University of Chemistry and Technology, 7, Sheremetievskiy Avenue, Ivanovo, 153000, Russia

E-mail: bsp@isuct.ru, i.astrakhantseva@mail.ru

**Abstract**. The article is devoted to the issues of using discrete simulation models for modeling some basic technological processes. In the scientific work, models in the form of multi-agent systems have been investigated, which allow us to consider a continuous environment as a set of interacting elements (agents), the behavior of which obeys local functions. The authors describe the basic techniques and general methodology for the development of deterministic agent-based models. The paper considers the use of multi-agent systems for modeling thermal conductivity, taking into account the nonlinearity of the process, in homogeneity of the material and the presence of volumetric heat sources of variable power in it. The obtained scientific results are in good agreement with the generally accepted classical approaches and do not contradict the provisions adopted in the theory of thermal phenomena.

## 1. Introduction

In recent years, for modeling many basic technological processes and systems, simulation approaches have begun to be used, which are based on a discrete consideration of objects and phenomena [1]. Among the methods that use discretization of time and space, agent-based modeling plays a significant role. It investigates the individual behavior of elementary subsystems (agents) that make up a large system [2; 3]. In this approach, the agent is considered as a kind of entity that has autonomy. It is believed that the global behavior of the system is a consequence of the individual behavior of agents [4].

There are a number of works showing that agent-based models can be a serious alternative to classical modeling methods based on the use of continual models in the form of systems of equations of various types. In particular, for the simulation of technological processes, cellular automata are successfully used - one of the varieties of agent-based models [5].

Practice has shown that the use of agent-based models with simple rules of behavior makes it possible to obtain adequate results when studying processes in inhomogeneous and anisotropic media, when solving problems for objects with complex geometry of boundaries, etc.

Another advantage of the agent-based approach is its object orientation, which allows the use of modern systems engineering techniques and greatly facilitates the development of software tools for simulating the phenomena under study.

## 2. Building an agent model
It is known that the basic technological processes in the construction, chemical, and energy industries are the processes of transfer of matter and energy (thermal, mechanical, etc.). It is obvious that all these phenomena occur in space. To use the agent-based approach to model such phenomena, dynamically interacting agents should be localized inside the object under study. Moreover, each localized agent will have neighbors with whom it contacts. Contacting agents interact with each other, which leads to a change in their states.

The behavior of agents can be described by an algorithm that allows one to determine their possible states; conditions that cause transitions between states; actions performed by agents [6]. The behavior algorithm of agents can be specified in the form of functional dependencies that explicitly relate the states of agents and the impact on them from their neighbors. As such dependences, it is advisable to apply the basic laws of a specific physical process. In particular, when modeling heat transfer, it is convenient to take its temperature as a parameter of the agent's state, an intense value that uniquely characterizes thermal phenomena. External influences, in this case, will be described by intense flux quantities, namely, heat fluxes between neighboring agents [7]. When studying diffusion phenomena, the state of an agent can be characterized by the concentration of the target component, which depends on the mass fluxes between neighboring agents. The change in the states of agents (their functioning) is carried out synchronously at discrete times.

Thus, a system of interconnected agents is introduced into consideration, the behavior of which will obey an individual algorithm based on the fundamental laws adopted for the modeled process [8].

As an example, consider the process of constructing an agent-based model of heat transfer by thermal conductivity. We will consider a two-dimensional problem.

Let a set of local agents located with a constant step $h$ (figure 1) are located inside the object under study.

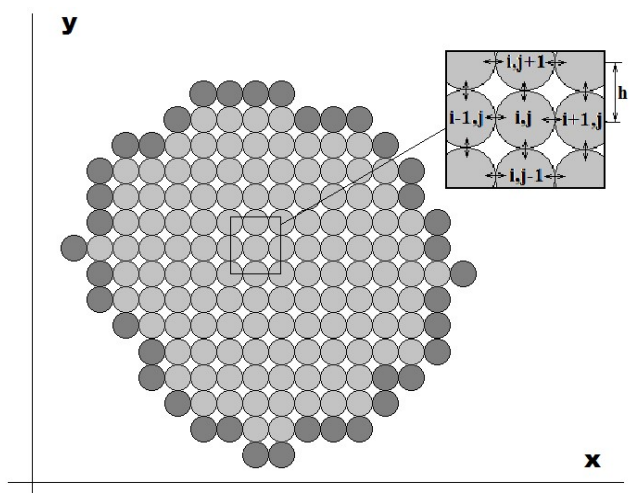

**Figure 1.** Localization of agents within the study area.

As shown above, the behavior of each agent is specified by an algorithm that determines the change in the agent's state as a result of its interaction with neighboring agents [9]. In this example, the working algorithm can be represented by functional dependence, which is not difficult to obtain using simple physical laws.

Since the interaction between agents occurs through the exchange of heat fluxes, it is possible to apply the Fourier law, according to which the vector of the heat flux is proportional to the temperature gradient [7].

Thus, we can write an expression for the heat flux to the agent in position $i,j$ from a neighboring agent at a discrete time moment $t_k$:

$$q_{i,j}(t_k) = \lambda_{i,j} \frac{\pm[T_{i,j}(t_k) - T_O(t_k)]}{h^2} \quad (1)$$

where $q$ is the specific power of the heat flow; $\lambda_{i,j}$ - coefficient of thermal conductivity of the agent material; $T_{i,j}(t_k)$ and $T_O(t_k)$ are the temperatures (states) of the considered and neighboring agents at time $t_k$; $h$ - coordinate step.

The specific sign in the numerator of the right-hand side of expression (1) is determined by the direction of the heat flux vector.

Since the agents, according to figure 1 are placed orthogonally, each of them will have four nearest neighbors. Consequently, four heat fluxes should be introduced into consideration, which determine the interaction of the agent with its neighbors.

The functional dependence for determining the temperature (state) of the agent at each discrete time step can be written as follows:

$$\frac{T_{i,j}(t_{k+1}) - T_{i,j}(t_k)}{\Delta t} = \frac{\sum_1^4 q_{i,j}(t_k)}{C_{i,j} \rho_{i,j}} \quad (2)$$

where $\Delta t$ is the time step; $C_{i,j}$ and $\rho_{i,j}$ are the heat capacity and density of the material of agent $i,j$ respectively.

This implies:

$$T_{i,j}(t_{k+1}) = T_{i,j}(t_k) + \frac{\Delta t}{C_{i,j} \rho_{i,j}} \sum_1^4 q_{i,j}(t_k) \quad (3)$$

Expression (3) determines the functioning of an individual agent in discrete time.

An important feature of expressions (1) - (3) is the fact that they contain specific physical characteristics of the material of an individual agent (density, heat capacity, thermal conductivity). This makes it possible to formulate an algorithm for the behavior of agents in the study of a process in heterogeneous environments.

The multi-agent approach assumes the presence in the system of agents whose functioning algorithms differ [10]. Such differences in behavior also take place in the study of the process of heat conduction. First, this applies to agents localized at different points in space. The equations obtained above are completely valid for agents that are inside the considered area and, accordingly, have four nearest neighbors. Agents located on the border (darker in figure 1) have fewer neighbors. However, the laws of their behavior are easy to obtain using the following reasoning. If the problem involves heat exchange with the external environment, then the fluxes at the boundary should be calculated using the heat transfer equations. If the boundary is isolated and there is no interaction with the surrounding space, then the number of terms in the sums included in expressions (2) and (3) should be reduced.

A separate type of agents of the system will be those whose state change is subject to external laws. These are, first, agents simulating the presence of heat sources or sinks in an object. In cases where a source is concentrated in one or several agents, then the law of changes in its state can describe its behavior either in time, or by the law of changes in the heat flux from it:

$$T_{m,n}(t_k) = \Psi(t_k) \quad (4)$$

or:

$$q_{m,n}(t_k) = F(t_k) \quad (5)$$

where $m, n$ are the coordinates of the agent; $T_{m,n}(t_k)$ is the agent temperature; $q_{m,n}(t_k)$ - heat flow; $\Psi(t_k)$ and $F(t_k)$ are given functions of discrete time (in the general case and material properties).

If the object has a distributed source (for example, a volumetric chemical reaction), then an additional term should be introduced into equation (3):

$$T_{i,j}(t_{k+1}) = T_{i,j}(t_k) + \frac{\Delta t}{C_{i,j}\rho_{i,j}}\left[\sum q_{i,j}(t_k) + \gamma(t_k)\right] \quad (6)$$

where $\gamma(t_k)$ is the specific power of the heat source (sink) at time $t_k$.

## 3. Results

Let's consider several examples of using agent-based models to simulate the heat conduction process.

A flat plate of regular shape was chosen as a model object, on which 1681 agents were placed with a step of 1 mm. The physical parameters of the material were taken as follows: density $\rho$ 1500 kg / m$^3$; specific heat $C$ 1000 J / (kg·K); thermal conductivity $\lambda$ 1.5 W / (m·K). When simulating the process, it was assumed that there was no heat transfer to the environment. The initial temperature of the plate was equal to 0 conditional degrees. The simulation step in time is 0.005 s. The results are presented below. The time is indicated in the upper right corner of the figures, the dimensions of the plate are presented on the horizontal $x$ and $y$ axes. The vertical axis shows the temperature in conventional units.

First, let us carry out simulation experiments under the assumption that the behavior of heat source agents obeys Eq. (4). Figure 2 the results of modeling the heating of the plate by a source located on one of its boundaries are presented. The source had a constant temperature of 20 degrees.

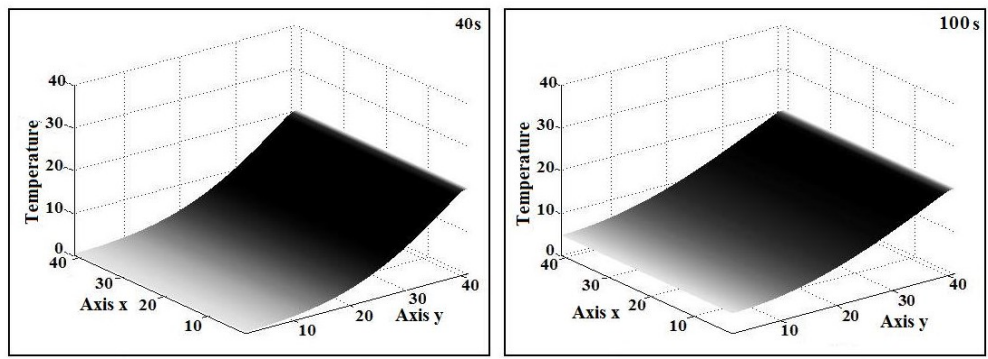

**Figure 2.** Results of modeling the heating of a plate by a linear heat source.

Figure 3 shows the change in the temperature of the plate when it is heated by a point heat source located in the center of the object. The source temperature was equal to 50 conditional degrees.

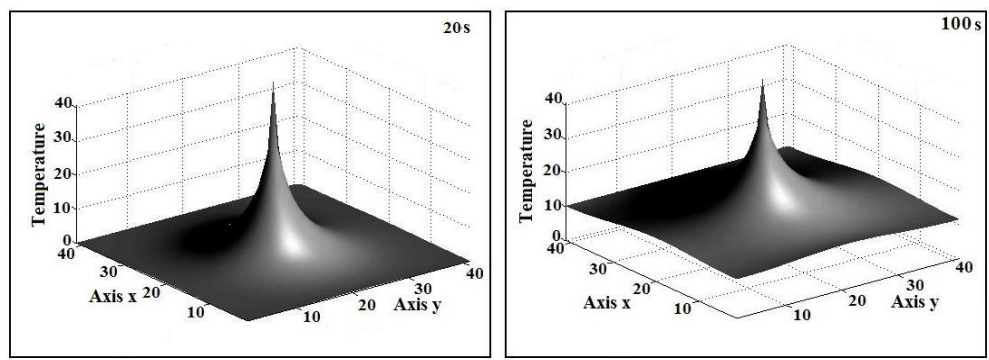

**Figure 3.** Results of modeling the heating of the plate by a point heat source.

Shown in figures 2 and 3 the results describe the process of gradual heating of the plate. In this case, its temperature asymptotically approaches a constant value, which is equal to the temperature of the source. This course of the process is fully consistent with existing ideas about the nature of molecular heat transfer.

In the simulation experiments, which are illustrated in figures 2 and 3, the functioning of the agents of the system-obeyed equations (1), (3), (4). These were trivial problems in a linear setting. Now we can move on to a more complex quasilinear problem. Let a volumetric heat source be located in the plate, and the functioning of the agents will obey Eq. (6). We also assume that the following relationship between the specific power of sources and temperature is valid:

$$\gamma(T) = kT \qquad (7)$$

where $k$ is a constant.

This formulation of the problem is typical for modeling heat transfer under conditions of heat release.

In the simulation, it was assumed that a heat pulse at the central point of the plate initiates combustion. The rest of the simulation parameters were taken similar to those described above. The simulation results are shown in figure 4.

Analysis of figure 4 shows that the plate is initially heated rather slowly. However, then the temperature rises sharply, which is typical for the combustion process.

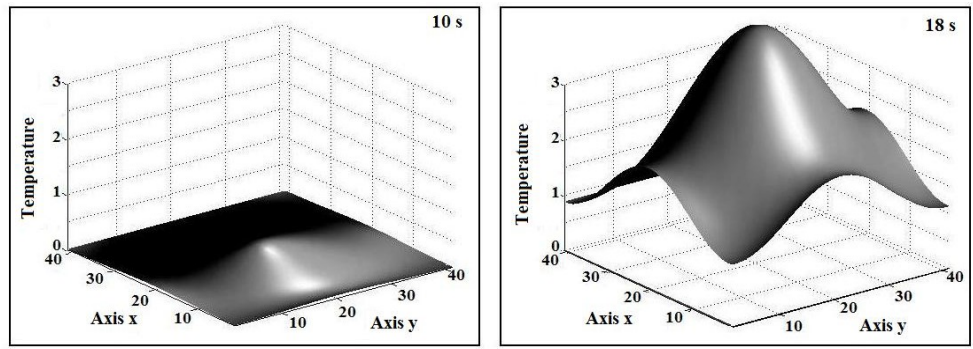

**Figure 4.** Results of modeling a quasilinear heat conduction process.

It was mentioned above that agent-based models make it easy to study processes in heterogeneous media. Next, we will consider heat transfer in a plate, which has areas with different physical characteristics. Let there be a section in the plate, the material of which has a significantly lower thermal conductivity compared to the bulk. Let us take these values $\lambda_0 = 1.5$ W / (m·K) and $\lambda_1 = 0.015$ W / (m·K). The rest of the simulation parameters will be the same.

Let us consider two cases with different locations of the anomalous area on the plate: in the first case, the area has a rectangular shape (figure 5), and in the second, it has a linear shape dividing the plate into two parts (figure 6).

Figures 5 and 6 show that the temperature of the anomalous zones is significantly lower than the general thermal field. In addition, the overall duration of the plate warm-up process is longer than in the previous example.

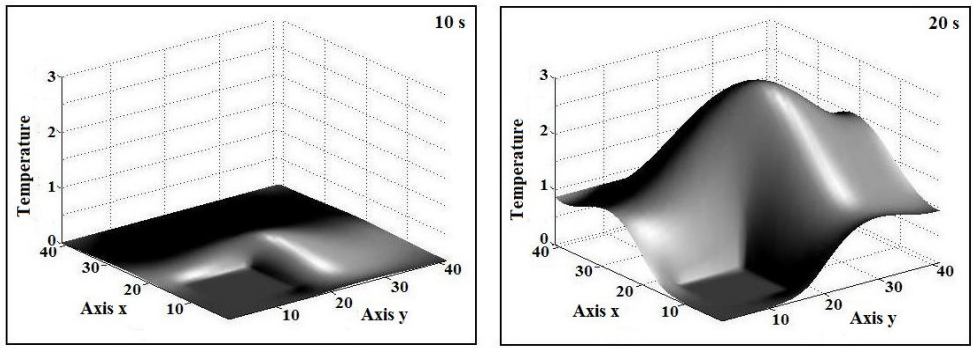

**Figure 5.** Results of modeling thermal conductivity in a plate with an anomalous zone.

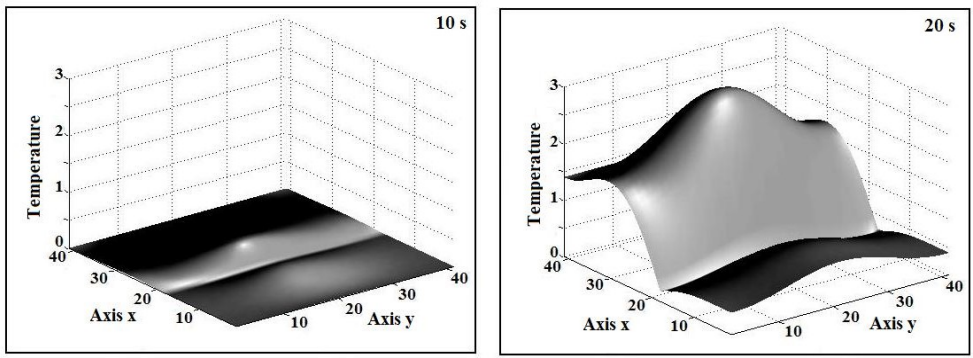

**Figure 6.** Results of modeling thermal conductivity in a plate with a heat-insulating section.

## 4. Discussion
All the results presented above are within the framework of existing concepts of the flow of heat transfer processes by the molecular mechanism. It should be noted that when the conditions of the problem were changed, the modeling computer program was subjected to the minimal adjustments. In general, the simulation algorithm remained unchanged. Only the dependencies underlying the algorithms for the behavior of agents and the characteristics of the material changed. When necessary, the coordinates of the location of anomalous areas were entered.

Summarizing what has been said; the following advantages of using agent-based models in modeling technological processes can be noted:

- The use of localized agents makes it possible to model objects with a complex shape of boundaries.
- Individual behavior of agents removes the problem of modeling when changing the properties of substances in space or time.
- The use of discrete agents facilitates the modeling and analysis of processes at the boundaries of regions in heterogeneous media.

The disadvantages of the presented approach include a significant need for computing resources, since an increase in the adequacy of a model requires an increase in the number of agents used in it. This, in turn, leads to an increase for computation. However, it should be noted that the growth of the capabilities of computer facilities and the use of parallel computing technology largely compensate for these disadvantages.

## 5. Conclusion
The discrete multi-agent approach allows you to create effective simulation models for studying spatially distributed technological processes. The agent-based approach allows you to consider a large and complex system as a set of interacting elements. That is, the initial overall task is broken down into several discrete small tasks. In this case, the actions of local elements shape the behavior of the system as a whole. The rapid growth of the technical capabilities of computer tools makes it possible to consider agent-based models as the basic apparatus for the development of simulation approaches to the study of complex technological processes and systems.